\documentclass[10pt, twocolumn, article]{revtex4}

\usepackage{amsmath}
\usepackage{amssymb}

\input epsf
\input rotate

\def\bea{\begin{eqnarray}}
\def\eea{\end{eqnarray}}
\def\ben{\begin{equation}}
\def\een{\end{equation}}
\def\benu{\begin{enumerate}}
\def\enu{\end{enumerate}}

\def\lsim {\ifmmode {\buildrel<\over\sim}}

\def\1var{(\bx_1...\bx\N)}

\def\half{\frac{1}{2}}

\def\b1{{\bf 1}}
\def\bx{{x}}

\def\N{_{\sss N}}

\def\sph_int{ {\int d^3 r}}

\def\infintd3r{ \int_{-\infty}^\infty d^3r\,}
\def\intd3r{ \int d^3r\,}

\def\laplace1d{\frac{d^2}{dx^2}}
\def\plaplace1d{\frac{d^2}{d{x'}^2}}

\def\padr2{\frac{\partial^2}{\partial r^2}}

\def\N{{\cal N}}
\def\a{{\alpha}}
\def\b{{\beta}}

\def\E{{\cal E}}

\def\Ro{{R_{\rm occ}}}
\def\Ri{{R_{\rm IP}}}
\def\Rd{{R_{\rm den}}}
\def\Rop{{R'_{\rm occ}}}

\numberwithin{equation}{section}

\begin{document}


\title{{Charge Transfer in Partition Theory}}
\author{Morrel H. Cohen}
\affiliation{Department of Physics and Astronomy, Rutgers
University, 126 Frelinghuysen Rd., Piscataway, NJ 08854, USA}
\affiliation{Department of Chemistry, Princeton University,
Washington Rd., Princeton, NJ 08544, USA}
\author{Adam Wasserman}
\affiliation{Department of Chemistry, Purdue
University, 560 Oval Drive, West Lafayette, IN 47907, USA}
\author{Roberto Car}
\affiliation{Department of Chemistry and Princeton Institute for the 
Science and Technology of Materials (PRISM), Princeton University, Princeton, NJ 08544, USA}
\author{Kieron Burke}
\affiliation{Department of Chemistry, University of California at
Irvine, 1102 Natural Sciences 2, Irvine, CA 92697, USA}

\begin{abstract}

The recently proposed Partition Theory (PT) ({\em J. Phys. Chem. A} {\bf 2007}, {\em 111}, 2229),  is illustrated on a simple one-dimensional model of a heteronuclear diatomic molecule. It is shown that a sharp definition for the charge of molecular fragments emerges from PT, and that the ensuing population analysis can be used to study how charge redistributes during dissociation and the implications of that redistribution for the dipole moment. Interpreting small differences between the isolated parts' ionization potentials as due to environmental inhomogeneities, we gain insight into how electron localization takes place in  H$_2^+$ as the molecule dissociates. Furthermore, by studying the preservation of the shapes of the parts as different parameters of the model are varied, we address the issue of transferability of the parts. We find good transferability within the chemically meaningful parameter regime, raising hopes that PT will prove useful in chemical applications.

\end{abstract}

\maketitle

\section{Introduction}

Consider a molecule of composition $AB$ with parts $A$ and $B$ having different ionization potentials when isolated. A long-standing problem is how to associate charges with each part as the parts are separated. At intermediate separations, one expects that, as bonding electrons would spend unequal time in the vicinity of each part, one would have to assign non-integer average numbers of electrons to each, numbers which become integers at infinite separation. Density-functional theory (DFT) is defined only for integer electron numbers as originally developed \cite{HK64, KS65}. If the dependences of the energy functionals of integer DFT on electron density were continued to densities containing noninteger electron numbers and applied to the separation of $AB$ into $A+B$, at infinite separation $A$ and $B$ would have unphysical noninteger electron numbers, as pointed out by Perdew et. al. (PPLB) \cite{PPLB82,P85}. Instead, PPLB argued that an ensemble generalization of ground-state DFT should be used for systems with noninteger electron number.

With how to treat noninteger electron number resolved by PPLB, the issue of how rigorously and systematically to decompose a system into its parts remains \cite{PAN05,GCGV06}. Two of the present authors have proposed an exact scheme, partition theory (PT) \cite{CW03,CW06,CW07}, based on the PPLB ensemble DFT. In \cite{CW07}, their PT was brought to full formal development and used for a reconstruction of chemical reactivity theory which eliminated the inconsistencies of earlier formulations and enriched them. Applying PT to the case introduced above, $AB\rightarrow A + B$, the parts would obviously be $A$ and $B$.

To illustrate the conceptual structure and physical content of PT, a very simple system was studied in \cite{CWB07}, a caricature of the hydrogen molecule consisting of two electrons moving independently in one dimension under the influence of two attractive delta-function potentials of equal strength, 1D$H_2$. In the present paper, a corresponding model of a heteronuclear diatomic molecule $AB$ is studied via PT (1D$AB$). The model once again consists of two electrons moving independently in one dimension under the influence of two attractive delta-function potentials of unequal strengths $-Z_A$ and $-Z_B$ with $Z_B<Z_A$. Since a one-electron $B$-atom would tend to donate its electron to the more electronegative $A$-atom when brought together, $A$ can be thought of as a Lewis acid and $B$ as a Lewis base.

In the limit $Z_A=Z_B=Z$, the model becomes that treated in \cite{CWB07}. Moreover, reducing the number of electrons from 2 to 1 requires little modification of that analytic theory, and its numerical results can be used to examine the dependence of charge transfer on $Z_A-Z_B$ and inter``nuclear'' separation. The resulting theory can also be used to explore how symmetry breaking localizes the single electron of $A_2^+$ when it is separated into $A$ and $A^+$, a subtler problem than the localization of both electrons on $A$ when $AB\rightarrow A^-+B^+$. 

In Section 2, the formalism developed for the 1D$H_2$ problem in ref.\cite{CWB07} is extended to the present 1D$AB$ problem. Numerical results are given in Section 3 for the dependence of the electron densities of the parts, for the charge transfer from $A$ to $B$, and for the dipole moment as functions of $Z_A$, $Z_B$, and the inter``nuclear'' separation $R$. Results are also given there for the partition potential of PT, and how it induces electronegativity equalization between the parts is discussed. The transferability of the properties of the atoms is discussed as well. The united-atom limit, of more academic interest than chemical relevancy, is discussed separately at the end of Section 3. Section 4 is devoted to the one-electron molecule 1D$AB^+$. The case $Z_A\downarrow Z_B$ is used to show how trivial symmetry breaking, $(Z_A-Z_B)<<Z_A$, is sufficient to localize the electron on $A$, illustrating how real $H_2^+$ separates into $H$ + $H^+$ because of small environmental perturbations. We conclude in Section 5 with a brief discussion of the significance of these very simple illustrations of the power and utility of PT for population analysis and for the transferability of fragments with their properties between different molecular contexts. Detailed derivations of all analytic results are presented in an Appendix.

\section{1D$AB$; independent electrons moving in unequal $\delta$-function potentials in one dimension}

\subsection{The Molecule}

In \cite{CWB07}, we considered an analogue of the $H_2$ molecule in which two electrons move independently in one dimension under the influence of two $\delta$-function potentials of equal strength $-Z$. In the present section, we consider the heteronuclear analogue 1D$AB$ in which the nuclear $\delta$-function of the acid $A$ is of strength $-Z_A$ and that of the base of strength $-Z_B$, with $Z_A>Z_B$. These ``nuclear charges'' are allowed to vary continuously. The ground-state wave functions $\psi_\a^0(x)$ and energies $E_\a^0$ of the isolated ``atoms'' $\a=A,B$ are (atomic units are used throughout):
\ben
\psi_\a^0(x)=\sqrt{Z_\a}\exp[-Z_\a|x|]~~,\label{e:3.1}\een
\ben E_\a^0=-Z_\a^2/2~~.\label{e:3.2}\een

The ground-state energy $E_M(N_M=1)$ of one electron moving independently in the two $\delta$-functions, that of strength $-Z_A$ at $x=-R/2$ and that of strength $-Z_B$ at $x=R/2$, is $-\kappa^2/2$, where
\ben
\left(\kappa-Z_A\right)\left(\kappa-Z_B\right)=e^{-2\kappa R}Z_AZ_B~~.
\label{e:3.3}
\een
The solutions of Eq.(\ref{e:3.3}) are plotted as a function of internuclear separation in Subsection A of the Appendix. From here on we are concerned only with the lowest-energy solution, denoted simply as $\kappa$, which corresponds to a bonding state that is doubly occupied when $N_M=2$.

The corresponding ground-state wave function, $\psi_M(x)$ is
\begin{eqnarray}
\nonumber\psi_M(x)&=&Ce^{\kappa(R/2+x)}~~~~~~~~~~,~~x<-R/2~~,\\
&=&De^{\kappa x}+Fe^{-\kappa x}~~,~~-R/2<x<R/2~~,\label{e:3.4}\\
\nonumber &=& G e^{\kappa(R/2-x)}~~~~~~~~~~,~~R/2<x~~,
\end{eqnarray}
where $C$, $D$, $F$, and $G$ are constants whose explicit expressions in terms of $Z_A$, $Z_B$, $R$, and $\kappa$ are given in subsection A of the Appendix (Eqs.\ref{e:3.5}, \ref{e:A2}).
In the combined-atom limit $R\downarrow 0$, $G=C=\kappa^{1/2}$, and $D$ and $F$ are irrelevant. In the limit $R\uparrow \infty$, $C=\kappa^{1/2}$, $D=e^{-\kappa R/2}\kappa^{1/2}$, and $G=F=0$ so that $\psi_M(x)$ is localized on $A$ only. 

\begin{figure}
\begin{center}
\epsfxsize=80mm \epsfbox{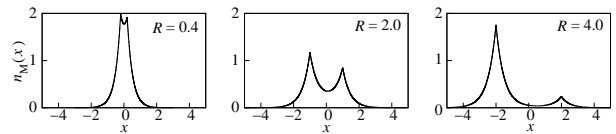}
\caption{Molecular density for three different internuclear separations: $R=0.4$ (left), $R=2.0$ (center), and $R=4.0$ (right). In this plot, $Z_A=1.02$, and $Z_B=0.98$.}\label{f:separation}
\end{center}
\end{figure}

The two-electron molecular density is 
\ben
n_M(x)=2|\psi_M(x)|^2~~;\label{e:3.7} 
\een
the total energy of the molecule $M$ is
\ben E_M(N_M=2)=2E_M(N_M=1)=-\kappa^2~~;\label{e:3.8} \een 
and the chemical potential of the molecule is
\ben \mu_M=E_M(2)-E_M(1)=-\kappa^2/2~~,\label{e:3.9} \een
all just as for the 1D$H_2$ case of ref.\cite{CWB07}.
Fig.\ref{f:separation} shows the dependence of $n_M(x)$ on internuclear distance $R$ for $Z_A=1.02$, $Z_B=0.98$. For these only slightly different values of $Z_A$ and $Z_B$ the transition from primarily ionic character (right panel) to mixed ionic-covalent character (center panel) occurs at relatively large bond length $R\sim 3.2$. The left panel of Fig.\ref{f:separation} shows a density belonging to the interesting but less chemically meaningful united-atom regime, a regime that we discuss separately in Subsection 2D because it allows us to draw conclusions regarding the limits of utility of Partition Theory.

\subsection{The parts}

Our task is to partition $n_M(x)$ into contributions from the two parts of $M$, fragments $A$ and $B$,
\ben n_M(x)=n_A(x)+n_B(x)~~,\label{e:3.10}\een 
with $n_{A,B}$ localized primarily around $-R/2$, $R/2$, respectively. Because $\psi_M$ and therefore $n_M$ is larger near $A$ than near $B$ (recall that $Z_A>Z_B$), the electron numbers of the fragments,
\ben N_{A,B}=\int dx n_{A,B}(x)~~,\label{e:3.11} \een
are unequal with $N_A > N_B$ and 
\ben N_A + N_B = N_M =2~~.\label{e:3.12} \een
$N_{A,B}$ are nonintegers in general so that Eq.(\ref{e:3.1}) implies that 
\ben N_A = 2- \nu~~,~~N_B = \nu~~, ~~0\leq \nu\leq 1~~,\label{e:3.13}\een
and the PPLB ensemble \cite{PPLB82} must be used for the fragments. In this use of the PPLB ensemble lies the main difference with Parr's atoms-in-molecules approach 
based on a minimum promotion energy criterion \cite{P84}. For $A$, a singly-occupied state of either spin occurs with probablility $\nu/2$, and a doubly-occupied state occurs with probability $1-\nu$ in the PPLB ensemble. For $B$, an unoccupied state occurs with probability $1-\nu$ and a singly-occupied state of either spin occurs with probability $\nu/2$.
The densities of the fragments are, accordingly,
\ben
n_A(x)=(2-\nu)\psi_A^2(x)~~,~~n_B(x)=\nu\psi_B^2(x)~~,\label{e:3.14} \een
where $\psi_\a(x)$ ($\a=A,B$) is a real one-electron wavefunction localized around $-R/2$, $\a = A$, or $+R/2$, $\a=B$. Analytic expressions for the $\psi_\a(x)$ are derived in subsections B through D of the Appendix. They can be written most simply as:
\ben
\psi_A(x)=(2-\nu)^{-1/2}\psi_M(x)\left(\cos\beta(x)+\sin\beta(x)\right)~~,
\label{e:psiA}
\een
\ben
\psi_B(x)=\nu^{-1/2}\psi_M(x)\left(\cos\beta(x)-\sin\beta(x)\right)~~,
\label{e:psiB}
\een
where $\beta(x)$ is an auxiliary polar-angle function that determines $\nu$ according to
\ben \nu = 1-\int dx \psi_M^2(x)\sin{2\b(x)}~~,\label{e:3.22}\een as shown in detail in Subsection B of the Appendix. The corresponding measure of electron transfer $q$ is
\ben
q=1-\nu=\int dx \psi_M^2(x)\sin{2\b(x)}
\label{e:2.15p}
\een

In Partition Theory, the wave
functions and densities of the parts are found by minimizing the sum of the
energies of the individual parts subject to the constraints that the electron
densities and numbers of the assembly of parts be identical to those of the
molecule $M$.  The expressions for $\psi_A(x)$ and $\psi_B(x)$ of Eqs.(\ref{e:psiA})-(\ref{e:psiB}) 
together with Eqs.(\ref{e:3.13})-(\ref{e:3.14}), guarantee that the constraints are met for 1D$AB$.
Only the determination of $\beta(x)$ remains, and the procedure for determining
it by minimizing the sum of the energies of the fragments is described in Subsections C and D of 
the Appendix.

\begin{figure}
\begin{center}
\epsfxsize=80mm \epsfbox{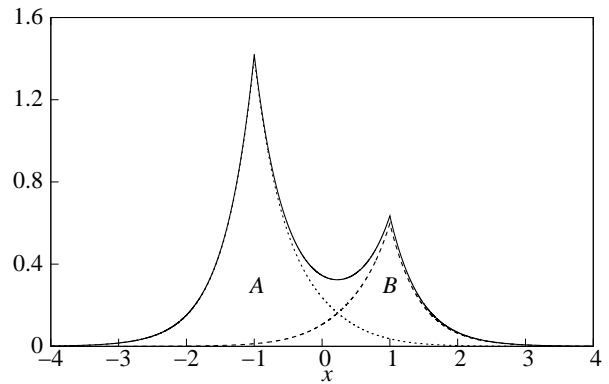}
\caption{Molecular density $n_M(x)$ (solid), and fragment
densities $n_A(x)$ (dotted) and $n_B(x)$ (dashed) for $Z_A=1.05$, $Z_B=0.95$ and $R=2$. For these values, $\nu=0.6$,
indicating substantial charge transfer even with only a 10\% difference between $Z_A$ and $Z_B$ and a small $R$.
}\label{f:density}
\end{center}
\end{figure}

Numerical evaluation of the quantities of interest, $\nu$, $\b(x)$, and $\psi_\a(x)$, is straightforward for representative $Z_\alpha$ and various $R$. For example, Fig. {\ref{f:density}} shows the fragments' atomic densities for $Z_A=1.05$, $Z_B=0.95$, and $R=2$. Interestingly, in spite of having high electron density in the bonding region around the molecule's center of mass, the fragment densities resemble true atomic densities. This is one of the most significant results of our work. In order to investigate the extent to which this is true, and quantify it, results for different parameter regimes are reported and discussed in the next Section. First, however, we discuss in the following subsection how the partition potential $v_P(x)$ of PT can be extracted from these results.

\subsection{The partition potential}

Within the framework of PT \cite{CW07}, the wavefunctions $\psi_\a(x)$ for the parts $\a=A,B$ satisfy the Schr\"{o}dinger equation
\ben
\left(H_\a+v_P(x)\right)\psi_\a(x)=\mu_M\psi_\a(x)~~,~~\a=A,B~~,
\label{e:3.41}
\een
where the Hamiltonians $H_\a$ for the parts are defined by Eq.({\ref{e:3.24}}). The partition potential $v_P(x)$ is {\em the same} for both parts, as is the eigenvalue $\mu_M$ corresponding to the molecular chemical potential, Eq.(\ref{e:3.9}). We construct $v_P(x)$ in subsection E of the Appendix, obtaining:
\begin{eqnarray}
 v_P(x)&=&-\half\frac{Z_A^2\psi_M^4(-R/2)\cos^2{2\b_A}}{\psi_M^4(x)}\theta(R/2-|x|)~~\label{e:3.46}\\
\nonumber &&~~~+\half\sum_\a v_\a\left(1-s_\a\tan\b_\a(1+\cos{2\b_\a})\right)~~,
\end{eqnarray}
where $\beta_\a$ ($\a=A,B)$ are constants that determine the upper and lower bounds of $\beta(x)$ (derived in subsection D of the Appendix), $s_\a=\pm 1$ (plus sign for part $A$, minus sign for part $B$), and $\theta(y)$($=0$ for $y<0$, $1$ for $y>0$) is the Heaviside step function. Numerical results for $v_P(x)$ are presented and discussed in the next Section.

\section{Illustrative Results}

For purposes of discussion, we define three critical parameters $\Ri$, $\Rd$, and $\Ro$ as the values of $R$ at which significant changes in the ionization potential, the density, and the occupation of a fragment, respectively, take place. Our definitions for $\Ri$ and $\Rd$ are immediately applicable to any diatomic molecule, and our definition for $\Ro$ could be easily generalized to be applicable to any diatomic molecule. We start by defining $\Ro$ in Subsection A in order to discuss population analysis and charge transfer, and continue in Subsection B with $\Rd$ and $\Ri$ to address the issue of transferability. 

\subsection{Population Analysis; Charge Transfer}
\begin{figure}
\begin{center}
\epsfxsize=80mm \epsfbox{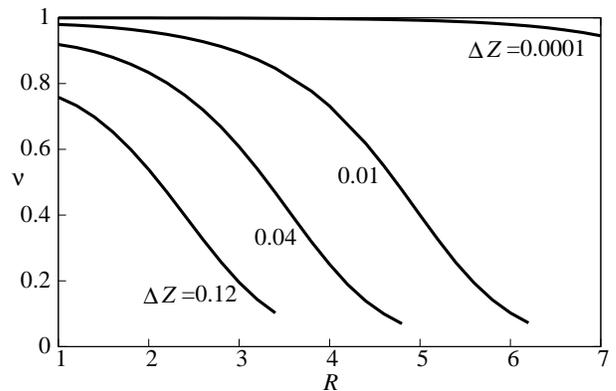}
\caption{Population of fragment $B$ (=$\nu$) for various values of $\Delta Z$, for $\bar{Z}=1$.}\label{f:nu_vs_a}
\end{center}
\end{figure}

 First, note that $q=1-\nu$ of an electron is transferred from $B$ to $A$. Figure {\ref{f:nu_vs_a}} shows how $\nu$ changes when $R$ is varied for 4 different values of $\Delta Z$ (and $\bar{Z}=1$) for $R>1$, and Fig.\ref{f:nu_vs_a_complete} does for smaller $R$ as well. Define $\Ro$ as the larger of the two values of $R$ for which $N_A=2N_B$ (and therefore $\nu=N/3$), a reasonable criterion for the ionic to mixed ionic-covalent crossover.  The solid curve in Fig.\ref{f:critical_a} displays the behavior of $\Ro$ vs. $\Delta Z$. We observe that $\Ro$ is a sensitive function of $\Delta Z$, especially as $\Delta Z\to 0$. (We come back to this point in the next paragraph and in Section 4 when discussing electron localization in H$_2^+$ as the molecule dissociates). This is a simple illustration of the utility of PT to quantify the degree of charge transfer taking place as a heteronuclear diatomic molecule is stretched out. That charge transfer occurs is readily seen from the sequence of molecular densities plotted in Fig.\ref{f:separation}, and PT provides an unambiguous way to characterize it.

This $R$-dependence of $\nu$ is interpretable as a sequence of crossovers in the electronic structure of the molecule $M$. At large $R$, $M$ is ionic with $\nu\downarrow 0$ and $q\uparrow 1$. As $R$ decreases, there is a crossover around $\Ro$ from that ionic to a mixed ionic-covalent state. As $R$ decreases further the covalent character of that mixed state increases. At still smaller $R$, there is another crossover to a combined-atom state around the maximum in $\nu$. Finally, at very small $R$, there is a rapid crossover back to the ionic state, an interesting fact that we discuss further in Subsection E. 

\begin{figure}
\begin{center}
\epsfxsize=80mm \epsfbox{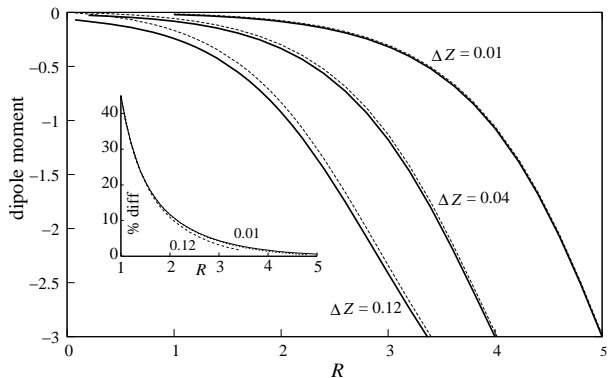}
\caption{Electronic dipoles as a function of inter``nuclear'' separation, both from the point-charge value of Eq.(\ref{e:dipole_PT}) (solid lines), and exactly (dashed), for 3 different values of $\Delta Z$. The inset shows the absolute percent difference. ($\bar{Z}=1$).}\label{f:dipoles}
\end{center} 
\end{figure}

\begin{figure}
\begin{center}
\epsfxsize=80mm \epsfbox{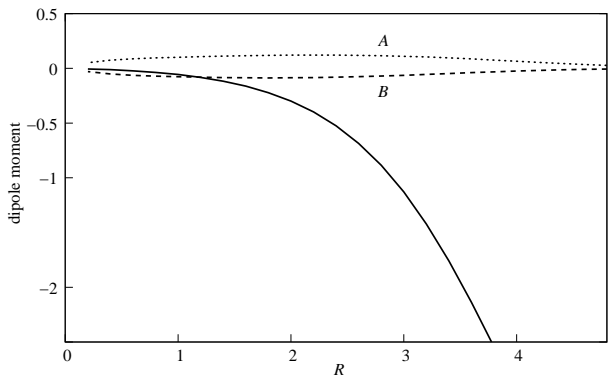}
\caption{Electronic dipole moment for the 1D$AB$ molecule (solid), and fragment dipole moments, for fixed $\Delta Z = 0.04$ and $\bar{Z}=1$.} \label{f:atomic_dipole}
\end{center} 
\end{figure}

It is also of interest to examine how well the product of transferred charge with the internuclear separation agrees
with the actual electronic component of the dipole moment of the molecule.  Figure \ref{f:dipoles} compares this point-charge dipole (PC)
\ben
d_{\rm PC}=-qR~~,
\label{e:dipole_PT}
\een
with the actual electronic part of the dipole moment $d=\int dx~xn_M(x)$ for various values of $\Delta Z$. If the
fragment densities were inversion-symmetric about their ``nuclei",
there would be exact agreement between $d_{\rm PC}$ and $d$. But there is in addition a dipole moment
on each fragment from the polarization of its electron cloud, $d_\a=\int dx~ xn_\alpha(x)$, $d=d_{\rm PC}+d_A+d_B$. Figure \ref{f:atomic_dipole} shows $d(R)$ along with the fragment dipoles $d_\a(R)$, $\alpha=A,B$, for fixed $\Delta Z$. Clearly, when $\Delta Z=0.04$ the fragment dipoles are small compared to the molecular dipole only for separations larger than $R\sim 3$. The estimate (\ref{e:dipole_PT}) shows excellent agreement whenever the dipole is significant, and only fails (error $> 20\%$) when the dipole is small (Fig.\ref{f:atomic_dipole}).
Note also from the inset of Figure \ref{f:dipoles} that the $R$-dependence of the percent difference changes little when $\Delta Z$ changes from 0.01 to 0.12. Figure \ref{f:atomic_dipole} shows that $d_A$ is of opposite sign to $d_B$ and $d_{\rm PC}$. This sign reversal arises from the distortion of $n_A$ by covalency, which increases it in the bonding region. However, in 1D$AB$ $d_A$ is not large enough in magnitude to overcome $d_B+d_{\rm PC}$ even at small $R$. If it were, $d$ would be of sign opposite to that expected from the electronegativity difference of $A$ and $B$. In real diatomic molecules containing small bases, such sign reversals are observed \cite{HMS93}, and partition theory promises a simple explanation.

\subsection{Transferability}
\begin{figure}
\begin{center}
\epsfxsize=80mm \epsfbox{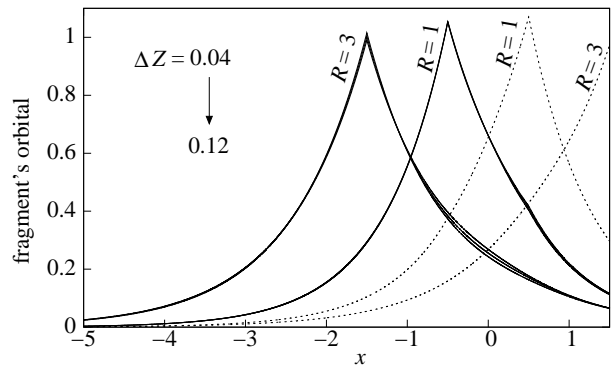}
\caption{Fragment orbitals $\psi_A(x)$ (solid) and $\psi_B(x)$ (dotted) for different values of $R$ (1 and 3). The fragment-$A$ orbitals are plotted for three different values of $\Delta Z$ (0.04, 0.08, and 0.12).  Orbitals corresponding to different values of $\Delta Z$ start to differ in the overlapping region only for large internuclear separations $R\gtrsim 3$. Differences due to changes in $\Delta Z$ are indistinguishable to the eye for $R\lesssim 3$.}\label{f:atoms_vs_a}
\end{center}
\end{figure}

\begin{figure}
\begin{center}
\epsfxsize=80mm \epsfbox{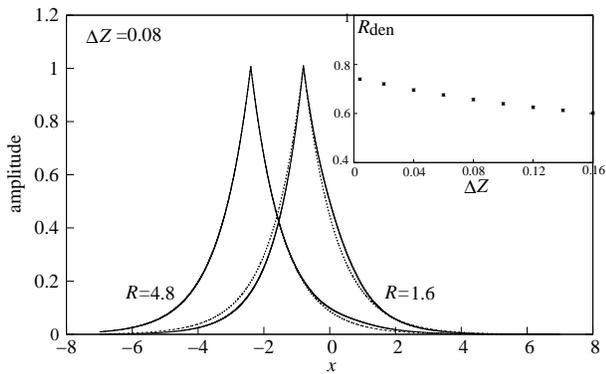}
\caption{Comparison of fragment orbital $\psi_A(x)$ (solid line) and isolated atomic orbital $\psi_A^0(x)$ (dotted) for $R=4.8$ and $R=1.6$. The inset shows the slow variation of $\Rd$ with $\Delta Z$ (see text).}\label{f:a_psi}
\end{center}
\end{figure}

\begin{figure}
\begin{center}
\epsfxsize=80mm \epsfbox{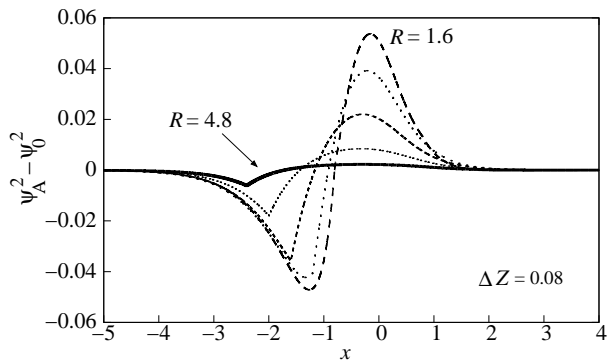}
\caption{Difference between the fragment density on $A$ and the isolated atomic density, $(\psi_A^0(x))^2-\psi_A^2(x)$, for different values of $R$ and fixed values of $\Delta Z=0.08$ and $\bar{Z}=1$. }\label{f:differences}
\end{center}
\end{figure}

\begin{figure}
\begin{center}
\epsfxsize=80mm \epsfbox{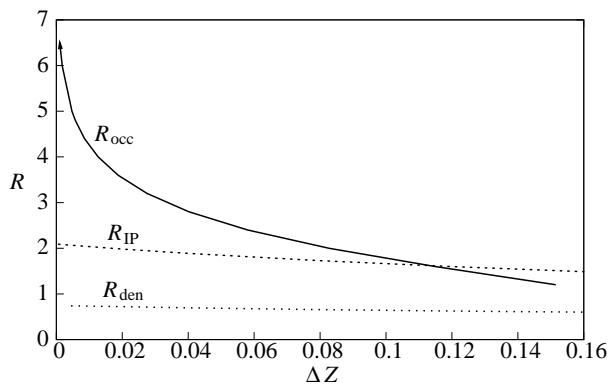}
\caption{Comparison of $\Ro$ (value of $R$ at which the occupation on $A$ is twice the occupation on $B$), $\Ri$ (at which the ionization potential of the molecule is 20\% larger than the ionization potential of an isolated $A$ atom), and $\Rd$ (at which the fragment densities change significantly as compared to the corresponding isolated-atom densities - see text). All curves with $\bar{Z}=1$.}\label{f:critical_a}
\end{center}
\end{figure}

\begin{figure}
\begin{center}
\epsfxsize=80mm \epsfbox{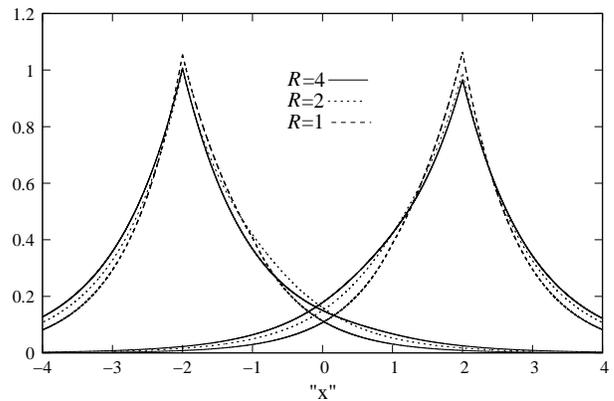}
\caption{Fragment orbitals $\psi_A(x)$ and $\psi_B(x)$ for $R=4$ (solid), $R=2$ (dotted), and $R=1$ (dashed), when $Z_A=1.03$ and $Z_B=0.97$. The A-fragment corresponding to $R=2$ was shifted to the left by 1a.u. and the B-fragment was shifted to the right by 1a.u. The A-fragment corresponding to $R=1$ was shifted to the left by 1.5a.u., and the B-fragment was shifted to the right by 1.5a.u.}\label{f:comparison}
\end{center}
\end{figure}

The mixed covalent to ionic crossover occurs at quite small $R$ for very small $\Delta Z$ when $\bar{Z}=1$. For the crossover to occur at an internuclear separation of about 1${\rm \AA}$, $\Delta Z$ need only be about 0.09.
The {\em shape} of each atom, however, is not as sensitive. Figure {\ref{f:atoms_vs_a}} shows $\psi_A(x)$ (solid) and $\psi_B(x)$ (dotted) for two different separations and three different values of $\Delta Z$ covering the same range as that of Fig.{\ref{f:nu_vs_a}}. It is apparent that orbitals corresponding to different values of $\Delta Z$ start differing significantly only close to the region where $\psi_A(x)$ and $\psi_B(x)$ overlap (the ``bonding'' region), and only for large internuclear separations $R\gtrsim 3$. Since at large internuclear separations the orbitals $\psi_\alpha(x)$ become equal to the isolated solutions $\psi_\alpha^0(x)$ of Eq.(\ref{e:3.1}), it is interesting to examine how and when the $\psi_\alpha(x)$ start departing from the $\psi_\alpha^0(x)$. Figure {\ref{f:a_psi}} shows that for $\Delta Z=0.08$, $\psi_A(x)$ is almost identical to $\psi_A^0(x)$ when $R=4.8$, differing only slightly in its right-hand tail, and that the shape is still preserved for chemically-relevant values like $R\sim 1.6$. In order to appreciate the differences more clearly, Fig.{\ref{f:differences}} displays the differences of the squares $D(x)\equiv \psi_A(x)^2-(\psi_A^0(x))^2$. We observe that the two orbitals depart appreciably when the spatial integral of the absolute value of this quantity reaches a value of $\int dx |D(x)| \sim 0.2$, so we define $\Rd$ as the corresponding separation. As shown in the inset of Fig.{\ref{f:a_psi}}, $\Rd$ remains almost constant at $\Rd\sim 0.66$, small compared to $\Ro$ in the corresponding range of $\Delta Z$. A similar behavior is observed for $\Ri$, which we define as the value of $R$ at which the ionization potential of the molecule begins to differ significantly [20\%] from the ionization potential of the most electronegative (isolated) atom. Figure {\ref{f:critical_a}} compares $\Ro$, $\Ri$ and $\Rd$. The preservation of the shape of the orbitals in a range of $\Delta Z$ where there is signficant charge transfer ($\Rd<<\Ro$) is a strong indication of the transferability of the fragments emerging from Partition Theory \cite{A00}. The feature of transferability can also be seen for fixed $\Delta Z$ by comparing fragments corresponding to different values of $R$. Figure {\ref{f:comparison}} shows such a comparison for $\Delta Z=0.06$ by overlapping the $\psi_A(x)$'s corresponding to different $R$'s (this requires shifting the origin of $A$-atoms to the left and $B$-atoms to the right). We note that even though the {\em size} of the fragments (controlled by $\nu$) changes substantially as $R$ varies from $1$ to $4$, their shape is quite insensitive to relatively large changes in $R$ with concomitantly large changes in $n_M(x)$ and $\mu_M$. For the value of $\Delta Z$ chosen in Fig.\ref{f:comparison}, $\nu$ is only about 0.2 when $R=4$, and yet the shape of the corresponding fragments only differ slightly from those obtained when $R=1$ and $\nu$ is close to 0.9.

\subsection{Partition Potentials}
\begin{figure}
\begin{center}
\epsfxsize=80mm \epsfbox{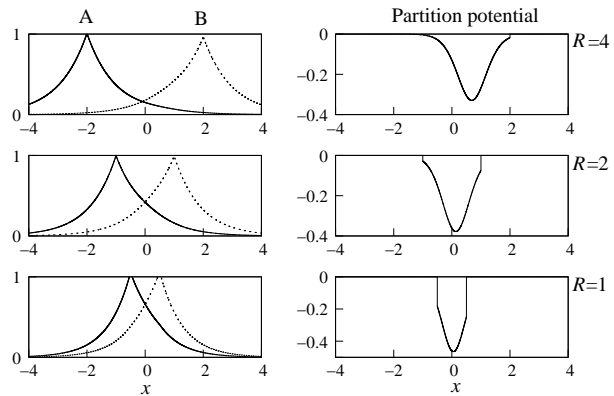}
\caption{Left panels: A and B fragments: $\psi_A(x)=\sqrt{n_A(x)/(2-\nu)}$ (solid) and $\psi_B(x)=\sqrt{n_B(x)/\nu}$ (dashed) for $R=4$ (upper), $R=2$ (center), and $R=1$ (bottom). Right panels: corresponding partition potentials ($\delta$-functions at $\pm R/2$ not shown). For all these plots, $Z_A=1.03$ and $Z_B=0.97$.}\label{f:partition_potential}
\end{center}
\end{figure}

\begin{figure}
\begin{center}
\epsfxsize=80mm \epsfbox{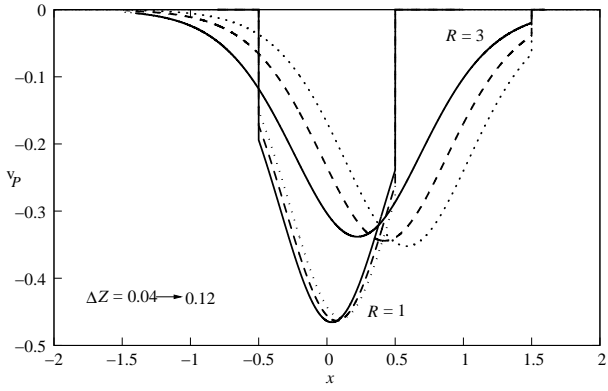}
\caption{Smooth part of the partition potential for 2 different values of $R$ and $\Delta Z = 0.04$ (solid), $\Delta Z = 0.08$ (dashed) and $\Delta Z=0.12$ (dotted). For all these curves, $\bar{Z}=1$.}\label{f:vR_vs_a}
\end{center}
\end{figure}

\begin{figure}
\begin{center}
\epsfxsize=80mm \epsfbox{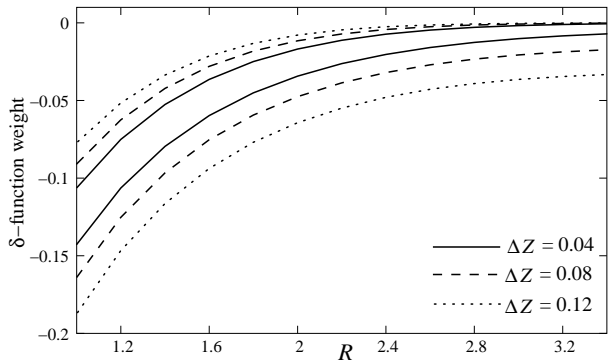}
\caption{Delta-function weights of the partition potential, for $\Delta Z=0.04$ (solid), $\Delta Z=0.08$ (dashed), and $\Delta Z=0.12$ (dotted). The 3 upper curves correspond to atom $A$, and the 3 bottom ones to atom $B$. For all these curves, $\bar{Z}=1$.}\label{f:weights}
\end{center}
\end{figure}

The corresponding partition potentials are shown in Fig. {\ref{f:partition_potential}}. Both parts $A$ and $B$ must have equal electronegativities, sharing the same HOMO eigenvalue, Eq.(\ref{e:3.41}), which must be equal to the overall chemical potential $\mu_M$. The partition potential ensures that this happens by acquiring a specific form, with an asymmetric negative well in between the fragments and two negative delta-functions at $\pm R/2$. In the limit of infinite separation, when $\kappa\to Z_A$, the external potential of ``nucleus'' $A$ does not require any correction in order to reach $\mu_M$, but the external potential of ``nucleus'' $B$ does. Accordingly, the delta-function component of $v_P$ vanishes on $A$ at infinite separation, but {\em not} on $B$ (fig.\ref{f:weights}). It is negative there, and has a magnitude smaller than $\Delta Z$ because the smooth negative well persists at large separations, getting further from $A$, and closer to $B$ (Figs.\ref{f:partition_potential}-\ref{f:vR_vs_a}), thereby contributing to the lowering of the eigen-energy of part $B$ towards $\mu_M$.  Figure {\ref{f:vR_vs_a}} shows a closer view of the dependence of the smooth part of the partition potential with $\Delta Z$ and with $R$, and Figure {\ref{f:weights}} illustrates the same dependences for the amplitudes of its delta-function components.

\subsection{United-atom limit}

\begin{figure}
\begin{center}
\epsfxsize=80mm \epsfbox{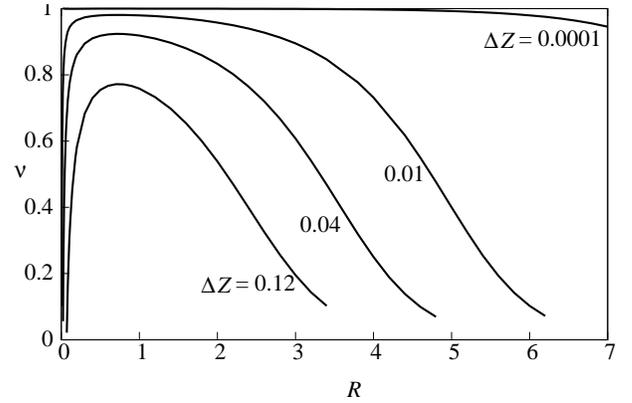}
\caption{Same as Fig.\ref{f:nu_vs_a} with added small-$R$ range to illustrate striking united-atom behavior of $\nu$ (see text).}\label{f:nu_vs_a_complete}
\end{center}
\end{figure}

In order to focus attention on the chemically relevant range of internuclear separations, we started Fig.\ref{f:nu_vs_a} at $R=1$. The small-$R$ behavior of $\nu$ can be seen in Fig.\ref{f:nu_vs_a_complete}. Perhaps counterintuitively, as $R$ decreases below a given (small) value, a rapid crossover takes place from a mixed ionic-covalent state to an essentially ionic state. In the 1D$H_2$ model of ref.\cite{CWB07}, this reversion to the ionic state is absent, the sequence of crossovers being atomic to covalent to combined-atom state. There is a singularity in the 1D$AB$ model at $\Delta Z=0$, $R=0$. If $\lim_{R,\Delta Z\downarrow 0}\left\{\frac{\Delta Z}{R}\right\}\downarrow 0$, the combined-atom state persists to $R=0$. If, however, $\lim_{R,\Delta Z\downarrow 0}\left\{\frac{R}{\Delta Z}\right\}\downarrow 0$, a reversion to the ionic state occurs. This counterintuitive feature illustrates an important limitation to the utility of partition theory. When $\kappa R$ becomes significantly less than unity as $R$ decreases, the $\psi_\alpha(x)$ overlap substantially, and the primary motivation of partition theory, decomposition of the electron density into distinct {\em localized} components, is frustrated.

The shape of the fragment densities also departs significantly in this limit from pure atomic densities. For example, for the same value of $\Delta Z = 0.08$ used in Fig.\ref{f:differences}, the maximum value of the difference between $\psi_A(x)^2$ and $\psi_A^0(x)^2$ when $R=0.4$ is about 6 times larger than the corresponding maximum when $R=1.6$.

\section{Electron localization in Dissociating H$_2^+$}

\begin{figure}
\begin{center}
\epsfxsize=80mm \epsfbox{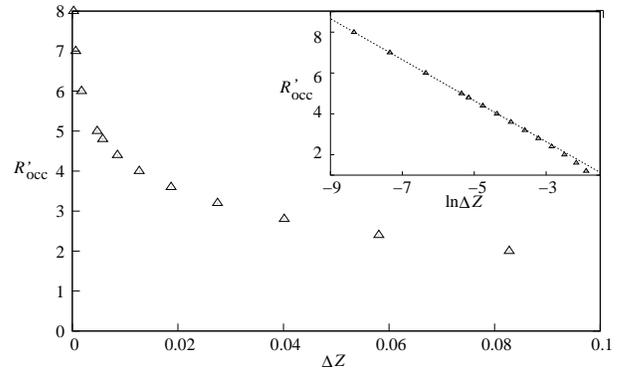}
\caption{Behavior of $\Rop$ for 1D$AB^+$ when $\bar{Z}=1$. The inset shows $\Rop$ vs. $\ln\Delta Z$. The line corresponds to the best linear fit in the range $-9\leq\ln\Delta Z\leq -5$ (slope: -0.185; intercept: -0.502, agreeing with the tight-binding formula derived in the Appendix, Eq.(\ref{e:tight_binding}))}\label{f:a_nu1}
\end{center}
\end{figure}

Even though the ground-state wavefunction of H$_2^+$ is symmetric, with 50\% of its amplitude on the right atom, and 50\% on the left, the slightest asymmetry due to environmental perturbations forces the electron to localize onto one of the two nuclei as the molecule dissociates. Since this symmetry breaking can now be studied experimentally via intense few-cycle laser pulses with controlled field evolution \cite{KSVK06}, there is resurgent interest in theoretical models to describe electron localization during molecular dissociation (see for example ref.\cite{KRG08} for a recent study of dissociation and ionization of small molecules steered by external noise). Our simple theory of the preceeding sections can be used as such a model, provided we interpret small differences between the magnitudes of $Z_A$ and $Z_B$ as due to the effect of an inhomogeneous environment. In fact, since we have dealt with 2 {\em non-interacting} electrons, only minor modifications of our results are needed in order to analyze the one-electron case, 1D$AB^+$ \cite{CD38}: the chemical potential is still identical to that given by Eq.(\ref{e:3.9}). The number constraint of Eq.(\ref{e:3.12}) is modified to $N_M'=1$ (we use primed symbols to represent 1-electron quantities to distinguish them from their 2-electron analogs), and Eq.(\ref{e:3.13}) goes to:
\ben
N_A'=1-\nu'~~,~~N_B'=\nu'~~,~~0\leq\nu'\leq\half~~.
\label{e:4.1}
\een
The densities of the atoms are:
\ben
n_A'(x)=(1-\nu')\psi_A^2(x)~~,~~n_B'(x)=\nu'\psi_B^2(x)~~,
\een
and following the same steps leading to Eq.(\ref{e:3.22}), we find
\ben
\nu'=\half\left[1-\int dx \psi_M^2(x)\sin{2\beta'(x)}\right]~~.\label{e:nu_1dAB+}
\een
Since the Euler equation for $\beta(x)$, Eq.(\ref{e:3.28}), as well as the boundary conditions, remain unchanged, $\beta'(x)=\beta(x)$ and the values that $\nu'$ takes as a function of $R$ and $\Delta Z$ are simply half of those calculated for 1D$AB$, $\nu'=\nu/2$. As before, we define $\Rop$ as the value of $R$ for which $N_A' = 2N_B'$, corresponding to $\nu'=1/3$ (from Eq.(\ref{e:4.1}) and $N_A'+N_B'=1$). The behavior of $\Rop$ as a function of $\Delta Z$ is then identical to that of $\Ro(\Delta Z)$. The inset of Fig.(\ref{f:a_nu1}) shows that for small $\Delta Z$, $\Ro$ is proportional to $\ln\Delta Z$. We conclude that the more inhomogeneous the environment (the larger $\Delta Z$), the earlier electron localization occurs along the dissociation pathway, with a logarithmic dependence in this case. As
$\Ro$ becomes very large when $\Delta Z$ becomes small, this striking relation between $\Ro$ and
$\Delta Z$  follows from a simple tight-binding argument outlined in the Appendix (Subsection F).



\section{Concluding Remarks}

We have applied Partition Theory \cite{CW07} to a simple model of a heteronuclear diatomic molecule. We found analytic expressions for the densities of the parts, the charge associated with each of the molecular fragments, and the partition potential that guarantees electronegativity equalization. Numerical calculations for various parameter regimes allow us to reach important conclusions: (1) $\Ro$ has a strikingly different bahavior than $\Ri$ and $\Rd$ as $\Delta Z\to 0$ (Fig.\ref{f:critical_a}). Since $\Ro$ measures the value of inter``nuclear'' separation at which significant charge transfer occurs, and $\Rd$ and $\Ri$ measure the value of $R$ at which signifcant change in the {\em shape} of the fragment wavefunctions takes place, we conclude that the fragments of Partition Theory, at least within this simple model, are to large extent transferrable. (2) Environmental fluctuations (modeled by small finite $\Delta Z$) localize the single electron of H$_2^+$ onto one of the two nuclei. As H$_2^+$ dissociates, the more inhomogeneous the environment, the earlier localization occurs along the dissociation pathway. The explicit results reported both here and in ref.\cite{CWB07} both illustrate important features of Partition Theory \cite{CW07} and support the proposed use of PT for a broad range of applications to real systems including the sharp definitions of parts of a larger system, population and charge-transfer analysis, and the examination of transferability.

 \renewcommand{\theequation}{A.\arabic{equation}}
  \setcounter{equation}{0}  
  \section*{APPENDIX}  

\subsection{The Molecule}

\begin{figure}
\begin{center}
\epsfxsize=80mm \epsfbox{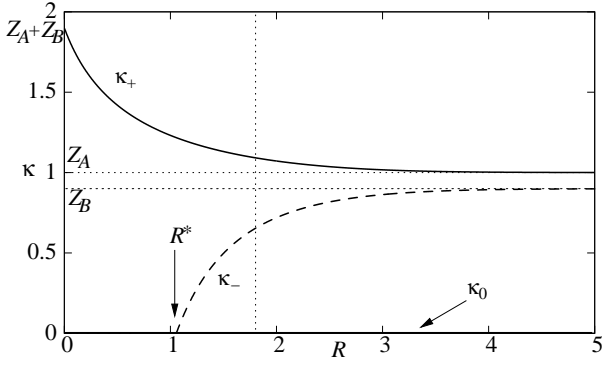}
\caption{$\kappa$ versus $R$ for $AB$ with $Z_A=1$ and $Z_B=0.9$. The upper curve shows the dependence of $\kappa$ for the doubly occupied orbital of the ground state which goes over from ionic on $A$ at infinite $R$ through mixed ionic-covalent to combined-atom at small $R$. The lower curve shows that of $\kappa$ for the empty orbital of the excited state which changes from atomic on $B$ at infinite $R$ to antibonding at intermediate $R$ and disappears for $R<R^*$; $\kappa_0=0$ is also a solution of Eq.(\ref{e:3.3}). The vertical dotted line indicates the separation $\Ri$ at which $\kappa$ begins to differ significantly (by 10\%) from $\kappa_+(\infty)$.
}\label{f:kappa_vs_a}
\end{center}
\end{figure}

Eq.(\ref{e:3.3}) has two solutions for all $R<(Z_A+Z_B)/2Z_AZ_B\equiv R^*$: one, $\kappa_+$, belonging to a bonding state doubly occupied when $N=2$, and another one, $\kappa_0=0$, belonging to a state at the bottom of the continuum. For all $R>R^*$ there is another solution, $\kappa_-$, corresponding to an unoccupied antibonding state. As $R\uparrow\infty$, $\kappa_+$ and $\kappa_-$ give rise to the two energies $E_{A,B}^0$ of Eq.(\ref{e:3.2}) corresponding to the two states $\psi_{A,B}^0(x)$ of Eq.(\ref{e:3.1}) of which the lower state $\psi_A^0(x)$ localized to the Lewis acid $A$ is doubly occupied, the Lewis base $B$ having donated its electron. As $R\downarrow 0$, $\kappa_+$ approaches $(Z_A+Z_B)$ and gives rise to a doubly occupied combined-atom state with the $Z_\a$ of Eqs.(\ref{e:3.1}) and (\ref{e:3.2}) replaced by $Z_A+Z_B$. The value of $\kappa_+$ decreases monotonically from $Z_A+Z_B$ at $R\downarrow 0$ with finite derivative at zero and vanishing derivative at infinity. $\kappa_-$ increases monotonically from zero at $R\downarrow R^*$ to $Z_B$ at $R\uparrow \infty$. $\kappa_0$ remains zero throughout (see Fig.{\ref{f:kappa_vs_a}}). In the text following Eq.(\ref{e:3.3}), we referred only to $\kappa_+$, dropping the subscript $+$ for notational simplicity. 

 The $R$-dependences of all three solutions of Eq.(\ref{e:3.3}) for $\kappa$ are shown in Fig.\ref{f:kappa_vs_a} for $Z_A=1, Z_B=0.9$. Note that $\kappa_+$ begins to differ significantly from $\kappa_+(\infty)$ only at separation  
$R<\Ri\sim 1.8$, less than 
relevant for real molecules, except for H$_2$. For example, for $Z_A=1$ and $Z_B=0.9$, one gets from Eqs.(\ref{e:3.3}) and (\ref{e:3.9}) an ionization energy equal to that of lithium hydride (LiH), for which the equilibrium bond distance is $R_0=3.05$ ($>>\Ri$).

Finally, the constants of Eq.(\ref{e:3.4}) are given in terms of $Z_A$, $Z_B$, $R$, and $\kappa$, by:
\ben
\left.
\begin{array}{lll}
D&=&e^{\kappa R/2}\left(1-Z_A/\kappa\right) C\\
F&=& e^{\kappa R/2}\left(Z_A/\kappa\right)C\\
G&=& e^{\kappa R}\left[(\kappa-Z_A)/Z_B\right]C
\end{array} 
\right\}~~,
\label{e:3.5}
\een
\begin{eqnarray}
C=&&(2\kappa)^{1/2}\left\{\frac{Z_A(\kappa-Z_A)}{Z_B(\kappa-Z_B)}\left(1+\frac{Z_B^2}{\kappa^2}\right)\right.\label{e:A2}\\
\nonumber&&\left.-\frac{e^{-2\kappa R}Z_A^2}{\kappa^2}+\frac{2Z_A}{\kappa}\left[1+2R(\kappa-Z_A)\right]\right\}^{-1/2}~~.
\end{eqnarray}

\subsection{The polar angle $\b(x)$}

Substituting Eqs.(\ref{e:3.14}) and (\ref{e:3.7}) into Eq.(\ref{e:3.10}) leads to
\ben (2-\nu)\psi_A^2(x)+\nu\psi_B^2(x)=2\psi_M^2(x) \label{e:3.15}~~.\een
Eq.(\ref{e:3.15}) can be rewritten as
\ben 
\chi_A^2(x)+\chi_B^2(x)   = \psi_M^2(x)~~,\label{e:3.16}\een
where
\ben
\chi_A(x)=(1-\nu/2)^{1/2}\psi_A(x)~~,~~\chi_B(x)=(\nu/2)^{1/2}\psi_B(x)~~,~~\label{e:3.17}\een
which permits us to take over the analytic procedures of ref.\cite{CWB07}. We first rotate $\chi_A(x)$ and $\chi_B(x)$ by $\pi/4$ in the 
function space in which they are defined, introducing 
\ben \chi_{\pm}(x)=\frac{1}{\sqrt{2}}\left(\chi_A(x)\pm\chi_B(x)\right)~~\label{e:3.18}\een
and leaving ``lengths'' within that space invariant so that 
\ben
\chi_+^2(x)+\chi_-^2(x)=\psi_M^2(x)~~.\label{e:3.19}\een 
Finally, we introduce the polar angle $\b=\b(x) $ in the function space,
\ben \chi_+(x)=\psi_M(x)\cos\b(x)~~,~~\chi_-(x)=\psi_M(x)\sin\b(x)~~,\label{e:3.20}\een
\ben \chi_{A,B}(x)=\frac{1}{\sqrt{2}}\psi_M(x)\left(\cos\b(x)\pm\sin\b(x)\right)~~.\label{e:3.21}\een
Because $\chi_{A,B}(x)$ are non-negative, $|\b|$ cannot exceed $\pi/4$. 

Inserting Eq.(\ref{e:3.21}) into Eq.(\ref{e:3.17}) and subtracting the squares of the two resulting equations leads to an expression for $\nu$, Eq.(\ref{e:3.22}):
\ben \nu = 1-\int dx \psi_M^2(x)\sin{2\b(x)}~~,\label{e:App_nu}\een
after integrating over $x$. Determination of the polar angle $\b(x)$ is thus sufficient for the determination of $\nu$ (charge transfer) and the electron population of the fragments (population analysis). For the symmetric 1D$H_2$ case, $\b(x)$ is odd and $\psi_M(x)$ even so that the integral in Eq.(\ref{e:3.22}) vanishes, yielding $\nu=1$ and equally populated fragments. In the present case, since $\psi_M(x)$ is normalized to unity and $|\sin 2\b(x)|\leq 1$ with the domain of positive $\b(x)$ weighted more heavily than that of negative $\b(x)$, the integral in Eq.(\ref{e:3.22}) lies in $(0,1)$, as must $\nu$ in accordance with Eq.(\ref{e:3.13}).

\subsection{The Euler equation for $\b(x)$}

In partition theory (PT) \cite{CW07}, a Hamiltonian is assigned to each part for each integer number of electrons entering into its PPLB ensemble. Since our ``electrons'' do not interact, it is sufficient to assign a one-electron Hamiltonian to each part,
\ben H_\a=\frac{p^2}{2}+v_\a(x)~~,~~v_{A,B}(x)=-Z_{A,B}\delta(x\pm R/2)~~.\label{e:3.24}\een
The PPLB energy functional of the collection of parts is then
\begin{eqnarray} 
\nonumber\E &=& (2-\nu)\left(\psi_A,H_A\psi_A\right)+\nu\left(\psi_B,H_B\psi_B\right)\\
&=&2\left[\left(\chi_A,H_A\chi_A\right)+\left(\chi_B,H_B\chi_B\right)\right]~~.\label{e:3.25}
\end{eqnarray}
Inserting the transformation (\ref{e:3.21}) and the definition (\ref{e:3.24}) of $H_\a$ into Eq.(\ref{e:3.25}) results in
\begin{eqnarray}
\nonumber
\E=\int dx \left\{\psi_M'^2(x)+\psi_M^2(x)\left[\b'(x)^2+v_A(x)+v_B(x)+\right.\right.\\
\left.\left.+(v_A(x)-v_B(x))\sin{2\b(x)}\right]\right\}~~~~~~~\label{e:3.26}
\end{eqnarray}
for the energy functional $\E$, now a functional only of $\b(x)$. In Eq.(\ref{e:3.26}) and in the following, primes indicate derivatives with respect to $x$.

Varying $\E$ with respect to $\b(x)$ yields
\begin{eqnarray}
\nonumber
\delta\E=2\int dx \psi_M^2(x)\left\{\b'(x)\delta\b'(x)+(v_A(x)-v_B(x))\right.\\
\times\left.\cos{2\b(x)}\delta\b(x)\right\}~~.\label{e:3.27}
\end{eqnarray}
The usual integration by parts leads to
\begin{eqnarray}
\nonumber
\delta\E=&&2\left\{\left.\psi_M^2(x)\b'(x)\delta\b(x)\right|_{x=-\infty}^{x=\infty}+\right.\\
&&+\left.\int dx\left[-\frac{d}{dx}\left(\psi_M^2(x)\frac{d\b(x)}{dx}\right)\right.\right.\\
\nonumber
&&+\left.\left.\psi_M^2(x)(v_A(x)-v_B(x))\cos{2\b(x)}\right]\delta\b(x)\right\}~~.\label{e:3.28}
\end{eqnarray}
The Euler equation
\begin{eqnarray}
\nonumber
-\frac{d}{dx}\left(\psi_M^2(x)\frac{d\b(x)}{dx}\right)+\psi_M^2(x)(v_A(x)-v_B(x))\\
\times\cos{2\b(x)}=0~~~~\label{e:3.29}
\end{eqnarray}
and the boundary conditions
\ben \left.\psi_M^2(x)\b'(x)\delta\b(x)\right|_{-\infty}^{\infty}=0\label{e:3.30}\een
result from imposing stationarity on $\E$. At first glance it might seem that Eq.(\ref{e:3.30}) is satisfied automatically since $\psi_M^2(x)\downarrow 0$ as $|x|\uparrow\infty$. However, unless the boundary condition 
\ben \b'(x)=0~~,~~|x|=\infty \label{e:3.31}\een is imposed, $\b'(x)$ diverges unacceptably as $\psi_M^{-2}(x)$ as $|x|\uparrow\infty$. 

\subsection{Solving for $\b(x)$}

Because the $v_\a(x)$ are $\delta$-function potentials, Eq.(\ref{e:3.29}) reduces to
\ben \frac{d}{dx}\left(\psi_M^2(x)\frac{d\b(x)}{dx}\right)=0~~,\label{e:3.32}\een 
with the additional boundary conditions
\ben \left.\begin{array}{l
l}\b(-\half R^+)=\b(-\half R^-)\equiv\b_A\\\b'(-\half R^-)-\b'(-\half R^+)=Z_A\cos{2\b_A}\end{array}\right\}~x=-\half R~,\label{e:3.33}
\een
\ben \left.\begin{array}{l
l}\b(\half R^-)=\b(\half R^+)\equiv\b_B\\\b'(\half R^-)-\b'(\half R^+)=-Z_B\cos{2\b_B}\end{array}\right\}~x=\half R~.\label{e:3.34}
\een
The general solution of Eq.(\ref{e:3.32}) is
\begin{eqnarray}
\frac{d\b(x)}{dx}&=&\frac{\a_1}{\psi_M^2(x)}~~,\label{e:3.35}\\
\b(x)&=&\int^x dx'\frac{\a_1}{\psi_M^2(x')}+\a_2~~,\label{e:3.36}
\end{eqnarray}
with the constants $\a_1$ and $\a_2$ taking on different values in the three domains $|x|>R/2$, $-\half R<x<\half R$. The condition (\ref{e:3.31}) implies that $\a_1$ and $\b'(x)$ vanish for $|x|>R/2$ so that
\ben 
\b(x)=\b_A~~,~~x<-R/2~~{\rm and}~~\b(x)=\b_B~~,~~x>R/2~~.\label{e:3.37}
\een
The conditions (\ref{e:3.33}) and (\ref{e:3.34}) imply that
\begin{eqnarray}
\nonumber
\a_1&=&-Z_A\psi_M^2(-R/2)\cos{2\b_A}\\
&=&-Z_B\psi_M^2(R/2)\cos{2\b_B}~~,~~|x|<R/2\label{e:3.38}
\end{eqnarray}
From Eq.(\ref{e:3.33}) $\a_2$ is $\b_A$ when the lower limit in Eq.(\ref{e:3.36}) is set at $-R/2$, yielding a second 
relation between $\b_A$ and $\b_B$,
\ben
\b_B-\b_A=\a_1\int_{-R/2}^{R/2}\frac{dx}{\psi_M^2(x)}~~,\label{e:3.39}
\een
Inserting the explicit form (\ref{e:3.4}) for $\psi_M(x)$ into the integral in Eq.(\ref{e:3.39}) results in
\ben
\int_{-R/2}^{R/2}\frac{dx}{\psi_M^2(x)}=\frac{1}{2\kappa D}\left[\frac{e^{\kappa R/2}}{\psi_M(-R/2)}-\frac{e^{-\kappa R/2}}{\psi_M(R/2)}\right]\label{e:3.40}
\een
Taken together, Eqs. (\ref{e:3.38}-\ref{e:3.40}) fix the values of $\b_{A,B}$, determining $\b(x)$ for $|x|>R/2$, and Eq.(\ref{e:3.36}) then determines $\b(x)$ for $|x|<R/2$. 

\subsection{The partition potential}

Since the $\chi_\a(x)$ are proportional to the $\psi_\a(x)$, Eq.(\ref{e:3.17}), they satisfy the same Schr\"{o}dinger Eqs.(\ref{e:3.41}). Summing over $\a$ and dividing by $\chi_A(x) +\chi_B(x)$ yields
\begin{eqnarray}
\nonumber
v_P(x)&=&\mu_M-\frac{1}{\chi_A(x)+\chi_B(x)}\frac{p^2}{2}(\chi_A(x)+\chi_B(x))+\\
&~&~~~-\frac{v_A(x)\chi_A(x)+v_B(x)\chi_B(x)}{\chi_A(x)+\chi_B(x)}~~.
\label{e:3.42}
\end{eqnarray}
Expressing $\chi_A(x)$ and $\chi_B(x)$ in terms of $\chi_+(x)$ and $\chi_-(x)$ via  Eq.(\ref{e:3.18}), and using Eq.(\ref{e:3.20}), results in
\begin{eqnarray}
\nonumber
v_P(x)&=&\mu_M+\frac{1}{2\psi_M(x)\cos\b(x)}\frac{d^2}{dx^2}(\psi_M(x)\cos\b(x))+\\
&~&~~~-\half\sum_\a v_\a(x)(1+s_\a\tan\b_\a)~~,
\label{e:3.43}
\end{eqnarray}
where $s_\a=1$ for $\a=A$ and $s_\a=-1$ for $\a=B$. The $\delta$-function character of $v_\a(x)$ and the definitions of $\b_A$ and $\b_B$ of Eqs.(\ref{e:3.32})-(\ref{e:3.33}) were also taken into account in arriving at Eq.(\ref{e:3.43}).
Since the molecular wavefunction $\psi_M(x)$ satisfies 
\ben
-\half\frac{d^2\psi_M(x)}{dx^2}+(v_A(x)+v_B(x))\psi_M(x)=\mu_M\psi_M(x)~~,
\label{e:3.44}
\een
Eq.(\ref{e:3.43}) can be transformed to
\begin{eqnarray}
\nonumber v_P(x)&=&-\half\left\{\tan\b(x)\left[\frac{2}{\psi_M(x)}\frac{d\psi_M(x)}{dx}\frac{d\b(x)}{dx}+\right.\right.\\
&~&~~~\left.\left.+\frac{d^2\b(x)}{dx^2}\right]+\left(\frac{d\b(x)}{dx}\right)^2\right\}\\
\nonumber
&&~~~+\half\sum_\a v_\a(x)(1-s_\a\tan\b_\a)~~.
\end{eqnarray}
Using the Euler Eq.(\ref{e:3.29}) for $\b(x)$, $v_P(x)$ can be further expressed as
\begin{eqnarray}
\nonumber
v_P(x)=-\frac{1}{2}\left(\frac{d\b(x)}{dx}\right)^2+\half\sum_\a v_\a(x)\\
~~~~~\times(1-s_\a\tan\b_\a(1+\cos{2\b_\a}))~.
\end{eqnarray}
Finally, by using Eqs.(\ref{e:3.36}) and (\ref{e:3.38}), we note that in the internuclear region $|x|<R/2$, $d\b(x)/dx$ is simply proportional to $\psi_M^{-2}(x)$, yielding
\begin{eqnarray}
 v_P(x)&=&-\half\frac{Z_A^2\psi_M^4(-R/2)\cos^2{2\b_A}}{\psi_M^4(x)}\theta(R/2-|x|)~~\label{e:3.46}\\
\nonumber &&~~~+\half\sum_\a v_\a\left(1-s_\a\tan\b_\a(1+\cos{2\b_\a})\right)~~,
\end{eqnarray}
where $\theta(y)\left(=0\right.$ for $y<0$, $1$ for $\left.y>0\right)$ is the Heaviside step function. Eq.(\ref{e:3.46}) correctly reduces to the partition potential of 1D$H_2$ \cite{CWB07} when $Z_A=Z_B$.

\subsection{Proof that $\Ro$ is proportional to $\ln\Delta Z$ for small $\Delta Z$}

We expect that $\Ro$, the nuclear separation at the crossover from covalent to ionic behavior, goes to infinity as $\Delta Z \downarrow 0$. There, the fragment wavefunctions must approach the free-atom wavefunctions, and the tight-binding LCAO must be a good approximation to the molecular orbital. We can thus write
\ben
\psi_M(x)=A\psi_A^0(x)+B\psi_B^0(x)~~
\een
where the $\psi_\a^0(x)$ are the orbitals of the isolated atoms, Eq.(\ref{e:3.1}). Taking matrix elements of the molecular Hamiltonian yields equations for $A$, $B$, and the molecular energy $E_M$:
\begin{subequations}
\begin{eqnarray}
\label{e:A.1a}\\
\nonumber
\left[E_A^0-E_M+v_B^{AA}\right]A+\left[(E_B^0-E_M)S_{AB}+v_A^{AB}\right]B=0\\
\nonumber
\left[(E_A^0-E_M)S_{AB}+v_B^{AA}\right]A+\left[E_B^0-E_M+v_A^{BB}\right]B=0 \\
\label{e:A.1b}
\end{eqnarray}
\end{subequations}
where the $E_\alpha^0$ are the energies of Eq.(\ref{e:3.2}), $v_{A/B}^{\alpha\beta}$ are matrix elements of the $A/B$ potentials of Eq.(\ref{e:3.24}), and $S_{AB}$ is the overlap $\left(\psi_A^0,\psi_B^0\right)$. 
In evaluating all quantities except $E_A-E_M$ and $E_B-E_M$, we can take the limit $\Delta Z\downarrow 0$. The result is
\begin{eqnarray}
\nonumber
S_{AB}=S_{BA}&\longrightarrow& S\sim ZRe^{-ZR}\\
\nonumber
v_B^{AA}\to v_A^{BB}&\longrightarrow& v_d\sim -Z^2e^{-2ZR}\\
\nonumber
v_A^{AB}\to v_B^{BA}&\longrightarrow& v_0\sim -Z^2e^{-ZR} 
\end{eqnarray}
The complete solution of Eqs.(\ref{e:A.1a})-(\ref{e:A.1b}) shows that $S$ enters into $E_M$, $A$, and $B$ as $1-S^2$ and $Sv_d$ which become exponentially small corrections as $\Delta Z\downarrow 0$, $R\uparrow \infty$. Similarly $v_d$ enters only in the combination $Sv_d$ and can be neglected as well. Thus the equations for $A$ and $B$ simplify to the classic bonding-antibonding equations
\begin{subequations}
 \begin{eqnarray}
  (E_A-E_M)A+v_0B&=&0\label{e:A.2a}\\
v_0A+(E_B-E_M)B&=&0\label{A:2b}
 \end{eqnarray}
\end{subequations}
The bonding eigenvalue is
\ben
E_M=\half(E_A+E_B)-\left\{\left[\half(E_A-E_B)\right]^2+v_0^2\right\}^{1/2}
\label{e:A.3}
\een
and 
\ben
\frac{B}{A}=\frac{E_M-E_A}{V_0}
\label{e:A.4}
\een
Inserting Eq.(\ref{e:A.3}) into Eq.(\ref{e:A.4}) and rearranging gives
\ben
\frac{B}{A}=\sqrt{1+\left(\frac{E_B-E_A}{2v_0}\right)^2}-\left|\frac{E_B-E_A}{2v_0}\right|~~.
\een
Now, since $E_B-E_A=Z\Delta Z$ and $v_0=-Z^2e^{-ZR}$ then 
\ben
\frac{B}{A}=\sqrt{1+\left(\frac{\Delta Z e^{ZR}}{2Z}\right)^2}-\frac{\Delta Z e^{ZR}}{2Z}\leq 1~~.
\een
Thus 
\begin{eqnarray}
\nonumber
\frac{B}{A}&=&f(y)\leq 1~~~,~~~y\equiv \frac{\Delta Z}{2Z} e^{ZR}\\
\nonumber
A&=&(1+f(y)^2)^{-1/2}
\end{eqnarray}
The molecular density is
\begin{eqnarray}
\nonumber
n_M(x)&=&N_M\psi_M(x)^2\\
\nonumber
&=&N_M \left[A^2\psi_A^2(x)^2+B^2\psi_B^2(x)+2AB\psi_A(x)\psi_B(x)\right]\label{e:A.5}\\
&=&\frac{N_M Z^2}{1+f(y)^2}(g_1(x)+g_2(x))~~,
\end{eqnarray}
where
\begin{eqnarray}
\nonumber
 g_1(x)&=&e^{-2Z|x|}+f^2(y)e^{-2Z|x-R|}\\
\nonumber
g_2(x)&=&2f(y)e^{-Z(|x|+|x-R|)}
\end{eqnarray}
It can be checked that $g_1(x)$ always exceeds $g_2(x)$ and becomes exponentially larger than $g_2(x)$ as $x$ departs from $\half R-\half(\ln f(y))/Z$. The cross term in Eq.(\ref{e:A.5}) can then be neglected and $n_M(x)$ is thus of the form obtained from Partition Theory: $n_M(x)=N_M A^2\psi_A(x)^2+N_M B^2\psi_B^2(x)$, so that $2A^2$ can be identified with $2-\nu$ and $2B^2$ can be identified with $\nu$, implying that 
\ben
f^2(y)=\frac{\nu}{N_M-\nu}~~.
\een
We have chosen $\nu=N_M/3$ to define $\Rop$. That implies that $f(y)=1/\sqrt{2}$ at $\Ro'$ for $N_M=1$ (or $f(y)=1/\sqrt{5}$ for $N_M = 2$). This leads to the observed behavior of $\Rop$ in Fig.(\ref{f:a_nu1}):
\ben
\Rop=\frac{1}{Z}\ln Z -\frac{1}{Z}\ln(\sqrt{2}\Delta Z)
\label{e:tight_binding}
\een

\end{document}